\documentstyle[11pt]{article}
\date{}
\newcommand{\eeq}{\end{eqnarray}}
\newcommand{\beq}{\begin{eqnarray}}

\newcommand{\bk}{{\bf k}}

\newcommand{\om}{{\omega}}

\newcommand{\bq}{{\bf q}}
\newcommand{\bv}{{\bf v}}
\newcommand{\bn}{{\bf n}}
\newcommand{\bp}{{\bf p}}
\newcommand{\bx}{{\bf x}}
\newcommand{\by}{{\bf y}}
\newcommand{\bK}{{\bf K}}

\newcommand{\cH}{{\cal H}}

\newcommand{\p}{{\partial}}

\def\con{{}_{\_\rule{-1pt}{0pt}\_}
\rule{-2pt}{0pt}\raise1.5pt\hbox{$\mid$}\hspace{2pt}}

\title{\bf Strong field limit of the Born-Infeld $p$-form
electrodynamics} 
\author{ Dariusz Chru\'sci\'nski\\
\it Institute of Physics, Nicholas Copernicus University\\
\it ul. Grudzi\c{a}dzka 5/7, 87-100 Toru\'n, Poland\\
\it e-mail: darch@phys.uni.torun.pl}

\begin{document}

\maketitle

\begin{abstract}

We study the strong field limit of $p$-form Born-Infeld theory. It turns out
that this limiting theory is a unique theory
displaying the full symmetry group of the underlying
canonical structure.
 Moreover, being a nonlinear theory, it possesses an
infinite hierarchy of conservation laws.

\vspace{0.5cm}



\end{abstract}

\vspace{1cm}

Preprint: hep-th/0005215 \\

Published in Phys. Rev. D {\bf 62} (2000) 105007/1-7.

\vspace{1cm}

\section{Introduction}
\setcounter{equation}{0}

The duality between strong and weak coupling regimes of the underlying
theory has played in recent years very prominent role (see e.g. review in
\cite{Olive}). In the present paper we study the weak-strong field limit
correspondence for the $p$-form Born-Infeld theory. Recently,
Born-Infeld nonlinear electrodynamics (BI) \cite{BI} has found a beautiful
applications in string theory and $p$-brane physics \cite{string}. The
motivation to study the corresponding $p$-form version of the BI theory
comes also from the string theory where one considers extended objects
($p$-branes) coupled to a $p$-form gauge potential \cite{T}.

In the weak field limit of $p$-form BI theory one obtains a linear
$p$-form Maxwell theory. The corresponding strong field limit is not so
well known. It was studied in \cite{IBB} for $p=1$ under the name Ultra
Born-Infeld theory (UBI). In the present paper we find the corresponding
$p$-form UBI and study its properties. It turns out that this theory being
nonlinear possesses very instructive features: it is invariant under the
full conformal group in (2$p$+2)-dimensional Minkowski space-time.
Us usual \cite{Deser}
the parity of $p$ plays a crucial role. However, in both cases (i.e. for
$p$ odd and even) the corresponding $p$-form UBI displays the full
canonical symmetry group of the underlying canonical structure,
 i.e. $SO(2,1)$ and $SO(1,1)\times Z_2$
symmetry for odd and even $p$ respectively.
Moreover, for odd $p$
it has an infinite hierarchy of conservation laws.
Therefore, the strong field limit of BI theory is even more symmetric than
the Maxwell theory which is also conformally invariant and being linear
has an infinite hierarchy of constants of motion.

\section{Born-Infeld $p$-form theory}
\setcounter{equation}{0}

\subsection{A general theory}

Consider a general nonlinear $p$-form electrodynamics defined in $D=2p+2$
dimensional Minkowski
space-time ${\cal M}^{2p+2}$ with the signature of the metric tensor
$(-,+,...,+)$.
The corresponding field tensor  $F=dA$ ($A$
denotes a $p$-form gauge potential) gives rise to the following
relativistic and gauge invariants:
\beq
S_p &=& - \frac{1}{2(p+1)!} F_{\mu_1...\mu_{p+1}} F^{\mu_1...\mu_{p+1}}\ ,\\
\label{Pp}
P_p &=& - \frac{1}{2(p+1)!} {F}_{\mu_1...\mu_{p+1}}
{\star F}^{\mu_1...\mu_{p+1}}\ ,
\eeq
 where the Hodge star
operation in ${\cal M}^{2p+2}$ is defined by:
\beq
{\star F}^{\mu_1...\mu_{p+1}} = \frac{1}{(p+1)!} \ \eta^{\mu_1...\mu_{p+1}
 \nu_1...\nu_{p+1}}\ F_{\nu_1...\nu_{p+1}}\
\eeq
and $\eta^{\mu_1\mu_2...\mu_{2p+2}}$ is the covariantly constant volume form
in the Minkowski space-time. The Hodge star satisfies $\star\star = (-1)^p$
which implies the crucial difference between $p$-form theories with different
parities of $p$.
 Having a Lagrangian
$L_p = L_p(S_p,P_p)$  one introduces a $G$-tensor
\beq   \label{const}
G^{\mu_1...\mu_{p+1}} = - (p+1)!\frac{\partial L_p}{\partial
F_{\mu_1...\mu_{p+1}}}\ .
\eeq
Eq. (\ref{const}) defines the constitutive relation for the underlying
$p$-forms electrodynamics. In the Maxwell theory one has $G(F)=F$ but in
the general case this relation may be highly nonlinear.

Now one may define the electric and magnetic intensities and inductions in the
obvious way:
\beq   \label{E}
E_{I} &=& F_{I0}\ ,\\   \label{B}
B_{I} &=& \frac{1}{(p+1)!}\
\epsilon_{IJk} F^{Jk}\ , \\   \label{D}
 D_{I} &=& G_{I0}\ ,\\         \label{H}
H_{I} &=& \frac{1}{(p+1)!} \
\epsilon_{IJk} G^{Jk}\ ,
\eeq
where we introduced a $p$-index $I=(i_1i_2...i_p)$ with $i_k=1,2,...,2p+1$.
$\epsilon_{i_1...i_pj_1...i_{p}k} = \epsilon_{IJk}$ denotes the
L\'evi-Civita tensor in $2p+1$
dimensional Euclidean space, i.e. a space-like hyperplane $\Sigma$ in the
Minkowski space-time.
Note, that $\epsilon^{i_1...i_{2p+1}} :=
\eta^{0i_1...i_{2p+1}}$.

  In terms
of $(E,B,D,H)$ the field equations $dF=0$ and $d\star G=0$ read:
\beq           \label{dB}
\partial_0 B^{I} &=& (-1)^p \frac{1}{p!}\
\epsilon^{IkJ}\
\p_k E_{J}\ ,\\    \label{dD}
\partial_0 D^{I} &=&  \frac{1}{p!}\
\epsilon^{IkJ}\
\p_k H_{J}\ .
\eeq
They are supplemented by the following  constraints (Gauss laws):
\beq   \label{Gauss}
\p_{i} B^{i...} =
\p_{i} D^{i...} = 0    \ .
\eeq
The dynamical properties of the $p$-form electrodynamics is fully
described by the energy-momentum tensor $T^{\mu\nu}$:
\beq
T^{\mu\nu} = \frac{1}{p!} \ F^{\mu I} G^{\nu}_{\ I} + g^{\mu\nu} L_p\ ,
\eeq
or in components:
\beq                              \label{T00}
T^{00} &=& \frac{1}{p!} \ E\cdot D - L_p\ ,\\
T^{0k} &=& (-1)^{p+1} (E \times H)^k\ ,\\
T^{k0} &=& (-1)^{p+1} (D \times B)^k\ ,\\
T^{kl} &=& - \frac{1}{(p-1)!} \left( E^{k...}D^l_{\ ...} +
H^{k...}B^l_{\ ...}  \right) + \delta^{kl} \left( \frac{1}{p!}
H\cdot B - L_p \right)\ ,
\eeq
where we introduced a convenient notation: $E\cdot D := E^I D_I$. Moreover,
\beq   \label{x}
(E \times H)^k = \frac{1}{(p!)^2} \epsilon^{kIJ} E_I H_J\ .
\eeq
Note, that
\beq
(E \times H)^k = (-1)^p (H \times E)^k\ .  \nonumber
\eeq
Now,
 observe that
dynamical equations (\ref{dB})-(\ref{dD}) have  already canonical form. The
Hamiltonian $\cH_p = T^{00}$ and
the corresponding Poisson bracket for the canonical variables
$(D^I,B^J)$ reads:
\begin{equation}   \label{Poisson}
\{ D^I(\bx), B^J(\by) \}_p = \epsilon^{IkJ}\, \p_k\, \delta^{(2p+1)}
(\bx-\by)\ ,
\end{equation}
(all other brackets vanish).

\subsection{Canonical symmetries}

\subsubsection{$p$ odd }

The $p$-form theory based on $L_p=L_p(S_p,P_p)$ is obviously relativistically
invariant. As is well known  in the Hamiltonian framework this
invariance is equivalent to the symmetry of the energy-momentum tensor.
Let us introduce the
following scalar quantities built out of the canonical variables $(D^I,B^J)$:
\beq   \label{alpha}
\alpha &=& \frac{1}{2p!}\ (D\cdot D + B\cdot B)\ ,\\  \label{beta}
\beta &=& \frac{1}{2p!}\ (D\cdot D - B\cdot B)\ ,\\   \label{gamma}
\gamma &=& \frac{1}{p!}\ D\cdot B\ .
\eeq
Now, the condition $T^{0k}=T^{k0}$ which is equivalent to
\begin{equation}  \label{s}
(E\times H)^k = (D\times B)^k\ ,
\end{equation}
implies the following equation for  $\cH_p$:
\begin{equation}        \label{O21}
(\p_\alpha \cH_p)^2 - (\p_\beta \cH_p)^2 - (\p_\gamma \cH_p)^2 =
1\ .
\end{equation}
Eq. (\ref{O21}) has a hyperbolic  $SO(2,1)$ symmetry. It turns out
that the group $SO(2,1)$  has a  natural representation on the
level of a canonical structure for a general $p$-form theory.

For $p$ odd the canonical structure defined in (\ref{Poisson}) is
invariant under:\\ 1) duality $SO(2)$ rotations: \beq   \label{I}
D^I &\rightarrow& D^I \cos\varphi - B^I\sin\varphi\ ,\nonumber\\
B^I &\rightarrow& D^I \sin\varphi + B^I\cos\varphi\ , \eeq 2)
hyperbolic $SO(1,1)$ rotations: \beq            \label{II} D^I
&\rightarrow& D^I \cosh\varphi + B^I\sinh\varphi\ ,\nonumber\\ B^I
&\rightarrow& D^I \sinh\varphi + B^I\cosh\varphi\ , \eeq 3)
$R^*$-scaling \beq                      \label{III} D^I
&\rightarrow& e^{\lambda}D^I\ ,\nonumber\\ B^I &\rightarrow&
e^{-\lambda}B^I\ . \eeq The easiest way to find the corresponding
generators for (\ref{I})-(\ref{III}) is to use a two potential
formulation \cite{Deser}, \cite{SS}. Let us introduce a $p$-form
potential $Z^I$ for $D^I$:
\begin{equation}
D^I = \epsilon^{IkJ}\, \p_k Z_J\ ,
\end{equation}
in analogy to
\begin{equation}
B^I = \epsilon^{IkJ}\, \p_k A_J\ ,
\end{equation}
where both $A^I$ and $Z^I$ are in the transverse gauge, i.e. $\p_i A^{i...} =
\p_i Z^{i...} = 0$. Defining $A_{(a)}^I = (A^I,Z^I)$ and $B^I_{(a)} =
(B^I,D^I)$ the corresponding generators may be written as follows:
\beq  \label{G1}
G_1 &=& \frac{1}{2p!} \int d^{2p+1}x \left( A_{(1)}\cdot B_{(1)} +
 A_{(2)}\cdot B_{(2)}\right) ,\\
G_2 &=& \frac{1}{2p!} \int d^{2p+1}x \left(A_{(1)}\cdot B_{(1)} -
A_{(2)}\cdot B_{(2)}\right) \ , \\
G_3 &=& \frac{1}{2p!} \int d^{2p+1}x \left( A_{(1)}\cdot B_{(2)}
+ A_{(2)}\cdot B_{(1)}\right) \ .
\eeq
For the Maxwell $p$-form theory one has $\cH_p^{Maxwell} = \alpha$ and
(\ref{O21}) is trivially satisfied. As is well know \cite{Deser} Maxwell
theory is invariant under (\ref{I}), i.e. $\{\cH_p^{Maxwell}, G_1\}=0$ and,
therefore, $G_1$ defines a constant of motion. Its physical meaning will be
clarified in subsection \ref{photons}.

\subsubsection{$p$ even }

For even $p$ the situation is very different. The invariant $P_p$ defined in
(\ref{Pp}) vanishes and therefore the general Hamiltonian $\cH_p$ does not
depend upon $\gamma$ defined in (\ref{gamma}). 
Eq. (\ref{O21}) reduces now to
\begin{equation}        \label{O11}
(\p_\alpha \cH_p)^2 - (\p_\beta \cH_p)^2 =
1\ .
\end{equation}
Eq. (\ref{O11}) displays the $SO(1,1)$ symmetry which is now
realized by (\ref{III}). Neither (\ref{I}) nor (\ref{II}) are
implementable as canonical transformations \cite{Deser} and
(\ref{III}) is generated by: 
\beq G_4 &=& \frac{1}{2p!} \int
d^{2p+1}x \left( A_{(1)}\cdot B_{(2)} - A_{(2)}\cdot
B_{(1)}\right) \ . \eeq However, in this case we have two
additional discrete $Z_2$ symmetries:
\begin{equation}  \label{Z1}
D^I \rightarrow B^I\ ,\ \ \ \ \ \ B^I \rightarrow D^I\ ,
\end{equation}
and
\begin{equation}  \label{Z2}
D^I \rightarrow - B^I\ ,\ \ \ \ \ \ B^I \rightarrow - D^I\ .
\end{equation}
Now, Maxwell theory is only $Z_2\times Z_2$-invariant, i.e. w.r.t. (\ref{Z1})
and
(\ref{Z2}).

\subsection{$p$--photons}   \label{photons}

For odd $p$ the canonical generator $G_1$ defined in (\ref{G1}) is constant
in time (in the Maxwell theory). To
find the physical interpretation of this quantity let us introduce a complex
notation:
\beq
F^I &=& D^I + i B^I\ ,\nonumber\\
G^I &=& E^I + i H^I\ .         \nonumber
\eeq
The dynamical equations (\ref{dB})-(\ref{dD}) rewritten in terms of
$(F^I,G^J)$ have the following form:
\begin{equation}
i \p_0 F^I = \frac{1}{p!}\, \epsilon^{IkJ}\, \p_k G_J\ .
\end{equation}
Moreover, let
\begin{equation}
V^I = Z^I + iA^I\ ,   \nonumber
\end{equation}
i.e. $F^I = \epsilon^{IkJ}\p_k V_J$.
Now, introduce a Fourier representation for $V(\bx)$:
\begin{equation}
\tilde{V}_I(\bk) = \int d^{2p+1}x\, e^{-i\bk\bx}\, V_I(\bx)\ .
\end{equation}
The transverse gauge $\p_i V^{i...}(\bx) =0$ implies $k_l
\tilde{V}^{l...}(\bk) =
0$. Let
$  \om^{(1)},\om^{(2)},...,\om^{(N_p)}  $
form an orthonormal  basis of $p$-forms on $2p$-dimensional plane
perpendicular to
$\bk$, i.e. $\om^{(i)}\cdot\om^{(j)} = \delta^{ij}$.
 The number $N_p$ reads
\beq
N_p = \left( \begin{array}{c} 2p\\ p \end{array} \right)\ , \nonumber
\eeq
and it is equal to the number of degrees of freedom for a $p$-form theory
\cite{moja}. Now, for any $\alpha \in \{ 1,2,...,N_p\}$ there exists exactly
one index $\alpha'$ such that
\begin{equation}
(\om^{(\alpha)} \times \om^{(\alpha')} )^l = \frac{k^l}{k}\ .
\end{equation}
Therefore, let us define a complex basis:
\begin{equation}
e^{(\alpha)} = \frac{1}{\sqrt{2}} (\om^{(\alpha)} + i \om^{(\alpha')} )\ ,
\end{equation}
where $\alpha$ runs from 1 to $N_p/2$.
Note, that the complex basis $e^{(\alpha)}$ satisfies:
\beq
\epsilon^{IlJ}\, k_l \, e^{(\alpha)}_{\ \ J} = - ik\, e^{(\alpha)I}
\eeq
and obviously: $\overline{e^{(\alpha)}}\cdot e^{(\beta)} =
\delta^{\alpha\beta}$ and $ {e^{(\alpha)}}\cdot e^{(\beta)} = 0$,
where
$\overline{a}$ denotes a complex conjugation of $a$.

Now, decomposing the Fourier transform of $V_I(\bx)$:
\beq
\tilde{V}_I(\bk) = \sum_{\alpha=1}^{N_p/2} \left( e^{(\alpha)}_{\ \ I}
f_+^{(\alpha)}(\bk) + \overline{e^{(\alpha)}}_{I} f_-^{(\alpha)}(\bk)
\right)\ ,
\eeq
and inserting into (\ref{G1}) one obtains:
\beq        \label{xx}
G_1 &=& \frac {1}{2p!} \int d^{2p+1}x \, V_I(\bx) \epsilon^{IkJ} \p_k
\overline{V_J(\bx)}
\nonumber  \\  &=&
\frac{1}{2p!(2\pi)^{2(2p+1)}}\ \int d^{2p+1}k\, k\,
 \sum_{\alpha=1}^{N_p/2} \left(
|f_+^{(\alpha)}(\bk)|^2 - |f_-^{(\alpha)}(\bk)|^2 \right)\ .
\eeq
Note, that the  energy of the Maxwell field reads:
\beq                  \label{xxx}
\cH_p &=& \frac {1}{2p!} \int d^{2p+1}x\, \left( D\cdot D + B\cdot B\right) =
\frac {1}{2p!} \int d^{2p+1}x\, F(\bx)\cdot \overline{F(\bx)}   \nonumber
\\  &=&
\frac {1}{2p!} \int d^{2p+1}x\, \epsilon^{IkJ}\,\p_k\, V_J(\bx)\,
\epsilon_{IjL}\, \p^j
\overline{V^L(\bx)}   \nonumber\\
&=& \frac {1}{2p!(2\pi)^{2(2p+1)}} \int d^{2p+1}k \, k^2
 \sum_{\alpha=1}^{N_p/2} \left(
|f_+^{(\alpha)}(\bk)|^2 + |f_-^{(\alpha)}(\bk)|^2 \right)\ . \eeq
Now,  $\cH_p$  measures the sum of intensities of all possible
polarizations. On the other hand $G_1$ measures the intensity of
right-handed polarizations minus the intensity of left-handed
polarizations. In the quantum theory of a $p$-form Maxwell field
they correspond to different polarization states of
``$p$-photons'' ($N_p/2$ left-handed defined by $e^{(\alpha)}$ and
$N_p/2$ right-handed defined by $\overline{e^{(\alpha)}}$).

Remarkably, (\ref{xxx}) is valid in the linear (Maxwell) case only
but (\ref{xx}) holds for any $p$-form theory and defines a
constant of motion for any duality invariant theory.
 Neither $G_2$ nor $G_3$
have any clear physical interpretation.

\subsection{Born-Infeld model}

The Born-Infeld $p$-form theory is defined by the following Lagrangian:
\begin{equation}    \label{LBI}
L^{(BI)}_p = \frac{b^2}{p!} \left( 1 - \sqrt{1 - 2b^{-2}p!S_p -
(b^{-2}p!P_p)^2 } \right) \  ,
\end{equation}
where $b$ denotes a generalized fundamental parameter of Born and Infeld
\cite{BI}.
In terms
of $E^I$ and $B^I$ our two basic invariants read:
\beq
S_p &=& \frac{1}{2p!}\, (E\cdot E - B\cdot B)\ ,\nonumber \\
P_p &=& 
\left\{ \begin{array}{ll} 
\frac{1}{p!}\,
E\cdot B\ , & \ \ \ \ \ \mbox{for\ odd}\ \ p\ ,\\
0 &  \ \ \ \ \ \mbox{for\ even}\ \ p\ .
\end{array}  \right.
\nonumber
\eeq
The corresponding $D^I$ and $H^I$ fields
read:
\beq   \label{DI}
D^I &=& \frac{1}{l_p} \left( E^I + b^{-2}P_p B^I \right)\ , \nonumber \\
H^I &=& \frac{1}{l_p} \left( B^I - b^{-2}P_p E^I \right)\ , \nonumber
\eeq
with $l_p =
  \sqrt{1 - 2b^{-2}p!S_p -
(b^{-2}p!P_p)^2 }$. From (\ref{T00}) one easily gets the corresponding
Hamiltonian:
\beq   \label{HBI}
\cH_p^{(BI)} =  \frac{b^2}{p!} \left(
\sqrt{ 1 + b^{-2}(D\cdot D + B\cdot B) + b^{-4}
\left[ (D\cdot D)(B\cdot B) - \epsilon_p\, (D\cdot B)^2 \right]
 } -1 \right) \ ,
\eeq
where
\beq
\epsilon_p = \left\{ \begin{array}{ll} 1 & \mbox{odd}\ p\\
0 & \mbox{even}\ p \end{array} \right. \ .  \nonumber
\eeq
Note, that for any $p$-forms $D$ and $B$
\beq
(p!)^2| D \times B|^2 =
(D\cdot D)(B\cdot B) + (-1)^p (D\cdot B)^2 \ .       \nonumber
\eeq
Therefore, for odd $p$
\beq   \label{HBI-odd}
\cH_p^{(BI)} =  \frac{b^2}{p!} \left(
\sqrt{ 1 + b^{-2}(D\cdot D + B\cdot B) + b^{-4}(p!)^2 |D\times B|^2
 } -1 \right) \ .
\eeq
 In terms of $(\alpha,\beta,\gamma)$ the BI Hamiltonian (\ref{HBI})
reads:
\beq   \label{HBI-alpha}
\cH_p^{(BI)} =  {b^2} \left(
\sqrt{ 1 + b^{-2}\alpha + b^{-4}(\alpha^2 - \beta^2 - \epsilon_p\, \gamma^2)
 } -1 \right) \ ,
\eeq
and satisfies (\ref{O21}) or (\ref{O11}) for odd or even $p$ respectively.

 It is easy to see that  $p$-form BI is
invariant under:
\begin{itemize}
\item dual $SO(2)$ rotations (\ref{I}) for $p$ odd,
\item $Z_2\times Z_2$ transformations (\ref{Z1}) and (\ref{Z2}) for $p$ even,
\end{itemize}
exactly as Maxwell theory.

\section{Strong field limit}
\setcounter{equation}{0}

The crucial property of the Born-Infeld model is that the magnitude of
 $E^I$ and $B^I$ fields is bounded by the value of the critical parameter
$b$, i.e. $|F^{\mu_1...\mu_{p+1}}|<b$. Therefore, we are not able
to study the strong field limit of this model in the Lagrangian
framework. Actually,  performing the $b\rightarrow 0$ limit in the
Born-Infeld Lagrangian (\ref{LBI}) one obtains $|P_p|$ which
defines a trivial theory. However, in the Hamiltonian framework
the $D^I$ field is not bounded and the question about the strong
field limit is well posed. The same property displays the
relativistic particle's dynamics: the particle's velocity is
always bounded $|\bv|<c$ contrary to its momentum $\bp$ (or energy
$E$) and
 the ultrarelativistic limit is defined by $|\bp|\gg mc$ (or $E\gg mc^2$). The
particle's Hamiltonian
\begin{equation}
H(\bq,\bp) = c\sqrt{\bp^2 + (mc)^2} + V(\bq)\ ,
\end{equation}
tends in the ultrarelativistic limit to
\begin{equation}   \label{p-H}
H^U(\bp) = c|\bp|
\end{equation}
which implies
\begin{equation}
\dot{\bp} = 0\ ,\ \ \ \ \ \bv = c\,\bn\ ,
\end{equation}
with $\bn = \bp/|\bp|$, i.e. one has a free evolution of photons.
By analogy we call the strong field limit of BI theory
the  {\it Ultra Born-Infeld} theory (UBI) \cite{IBB}.

\subsection{$p$ odd }

Performing $b\rightarrow 0$ limit in (\ref{HBI}) one gets
\begin{equation}   \label{HUBI}
\cH_p^{(UBI)} = |D \times B|\ .
\end{equation}
 Therefore, the constitutive relations are given by:
\beq   \label{EI}
E_I &=& p!\ \frac{\delta \cH_p^{(UBI)}}{\delta D^I} =
\epsilon_{kIJ}n^kB^J \ ,\\   \label{HI}
H_I &=& p!\ \frac{\delta \cH_p^{(UBI)}}{\delta B^I} =
- \epsilon_{kIJ}n^kD^J \ ,
\eeq
where $n^k$ stands for the unit (2p+1)-vector in the direction of the
generalized Poynting vector:
\beq    \label{nk}
n^k = \frac{(D\times B)^k}{|D \times B|}\ .
\eeq
The dynamical equations (\ref{dB})-(\ref{dD}) have the following
form:
\beq    \label{d1}
\p_0 B^I &=& - \delta^{\ kI}_{[jJ]} \, \p_k (n^j B^J)\ ,\\   \label{d2}
\p_0 D^I &=& - \delta^{\ kI}_{[jJ]} \, \p_k (n^j D^J)\ ,
\eeq
with $\delta^{ij...}_{kl...} = \delta^i_k\delta^j_l...$.
The remarkable property of $p$-form UBI is a structure of
 the energy-momentum tensor. One easily finds:
\beq             \label{t1}
T^{0k} &=& T^{k0}\ =\ {\cH}_p^{(UBI)}\, n^k\ ,\\   \label{t2}
T^{kl} &=& {\cH}^{(UBI)}_p\, n^kn^l\ .
\eeq
These relations were already derived by Bia{\l}ynicki-Birula \cite{IBB} for
$p=1$. But they hold for any odd $p$.
In terms of $(\alpha,\beta,\gamma)$ the UBI Hamiltonian (\ref{HUBI}) reads:
\beq   \label{HBI-1}
\cH_p^{(UBI)} =
\sqrt{ \alpha^2 - \beta^2 -  \gamma^2
 }  \ ,
\eeq
and displays the full $SO(2,1)$ symmetry of (\ref{O21}),
i.e.
$p$-form UBI is invariant under:
 (\ref{I}), (\ref{II}) and (\ref{III}).

Moreover, for any $p$ the trace of $T^{\mu\nu}$ for UBI vanishes and,
therefore,
UBI is invariant under the  conformal group in ${\cal M}^{2p+2}$. This
last property follows from the fact that UBI does not contain a
dimensional parameter.

Note, that after performing the Legendre transformation the UBI Lagrangian
vanishes:
$L_p^{(UBI)} = D\cdot E - \cH^{(UBI)}_p \equiv 0$. It does not mean, however,
that the theory is trivial (we already know that it is not). Vanishing of
$L_p^{(UBI)}$ denotes the presence of Lagrangian constraints. It is easy to
see that the constitutive relations imply $S_p=0$ and $P_p=0$.
Therefore, the UBI action has the following form:
\begin{equation}  \label{W}
W^{(UBI)}_p = \int d^{2p+2}x \left( \Lambda_1 S_p + \Lambda_2
P_p
\right) \ ,
\end{equation}
where $\Lambda_1$ and $\Lambda_2$ are Lagrange multipliers. Now, (\ref{W})
implies
\beq
D^I = \Lambda_1 E^I + \Lambda_2 B^I\ ,  \nonumber
\eeq
and hence $|D\times B| = \Lambda_1 B\cdot B$  (since $E\cdot E =
B\cdot B$),
 which gives $\Lambda_1 = |D\times B|/ B\cdot B$. Moreover, $D\cdot B =
\Lambda_2 B\cdot B$ and hence $\Lambda_2 = D\cdot B/ B\cdot B$. Inserting
\beq
E^I &=& \frac{1}{\Lambda_1} D^I - \frac{\Lambda_2}{\Lambda_1} B^I\ ,
\nonumber     \\
H^I &=& \frac{\Lambda_1^2 + \Lambda_2^2}{\Lambda_1^2} B^I -
\frac{ \Lambda_2}{\Lambda_1}  D^I\
,\nonumber
\eeq
into field eqs. (\ref{dB})--(\ref{dB}) one gets (\ref{d1})--(\ref{d2}).

\subsection{$p$ even }

Now, due to (\ref{HBI})
the UBI Hamiltonian simplifies to
\beq   \label{UBI-even}
\cH^{(UBI)}_p = \frac{1}{p!}\ |D||B|  \ ,
\eeq
giving rise to the following
 constitutive relations:
\beq   \label{c1}
E^I &=&  \frac{|B|}{|D|}\, D^I\ ,\\  \label{c2}
H^I &=&   \frac{|D|}{|B|}\, B^I\ .
\eeq
The corresponding stress tensor reads:
\begin{equation}   \label{T-even}
T^{kl} = \cH^{(UBI)}_p \left( \delta^{kl} - p \left( |D|^{-2} D^{k...}D^l_{\
...} + |B|^{-2} B^{k...}B^l_{\ ...} \right) \right)\ .
\end{equation}
Now, dynamical equations
(\ref{dB})-(\ref{dD}) read:
\beq           \label{d1e}
\partial_0 B^{I} &=&  \frac{1}{p!}\
\epsilon^{IkJ}\
\p_k \left( \frac{|B|}{|D|}\,D_{J}\right)\ ,\\    \label{d2e}
\partial_0 D^{I} &=&  \frac{1}{p!}\
\epsilon^{IkJ}\ \p_k \left( \frac{|D|}{|B|}\,B_{J}\right)\ . \eeq
Note, that (\ref{UBI-even}) rewritten in terms of $(\alpha,\beta)$
has $SO(1,1)$--invariant form: \beq \cH^{(UBI)}_p = \sqrt{\alpha^2
- \beta^2}\ . \eeq The theory displays the full symmetry of the
canonical structure, i.e. it is invariant under (\ref{III})
generated by $G_4$ and under $Z_2\times Z_2$ defined in
(\ref{Z1})--(\ref{Z2}). Obviously, $T^\mu_{\ \mu}=0$ and the
theory is conformally invariant. The Lagrangian structure may be
derived from the following action
\begin{equation}   \label{W1}
W^{(UBI)}_p = \int d^{2p+2}x\,  \Lambda\, S_p\ ,
\end{equation}
($\Lambda$ -- Lagrange multiplier) in analogy to (\ref{W}).

Note, that for Cauchy data satisfying
\beq   \label{DB}
D\cdot B =0\ .
\eeq
one has
\beq
|D \times B| = \frac{1}{p!}\, |D||B|\ ,
\eeq
and  the Hamiltonian (\ref{UBI-even}) has the same form as in
(\ref{HUBI}). The constitutive relations (\ref{c1})--(\ref{c2}) are
equivalent to:
\beq   \label{EIe}
E_I &=&  \epsilon_{kIJ}n^kB^J \ ,\\   \label{HIe}
H_I &=&  \epsilon_{kIJ}n^kD^J \ ,
\eeq
with $n_k$ defined in (\ref{nk}) and the stress tensor
(\ref{T-even}) may be rewritten as in (\ref{t2}).

\section{Fluid dynamics and new constants of motion}   \label{FLUID}
\setcounter{equation}{0}

In this section we generalize the observation made  in
\cite{IBB} for any odd  $p$. Observe that due to
(\ref{t1})-(\ref{t2}), $T^{\mu\nu}$ may be written in the following form:
\begin{equation}
T^{\mu\nu} = \cH^{(UBI)}_p\, U^\mu U^\nu\ ,
\end{equation}
where the (2$p$+2)-velocity
$U^\mu = (1,n^k)$ satisfies $U^\mu U_\mu=0$ (for even $p$ it holds for the
Cauchy data satisfying
(\ref{DB})).  Such a theory describes a dust of
particles moving with the speed of light in ${\cal M}^{2p+2}$ --
``$p$-photons''. It is easy to show that both the continuity equation
\begin{equation}  \label{c}
\p_\mu (\cH_p^{(UBI)}\, U^\mu) =0\ ,
\end{equation}
and the Euler equation
\begin{equation}           \label{Euler}
U^\nu\p_\nu U^\mu = 0
\end{equation}
are satisfied. Moreover, one easily proves that due to (\ref{c}) and
(\ref{Euler}) the following infinite set of continuity equations hold:
\begin{equation}
\p_\mu\, \left( \cH^{(UBI)}_p\, U^\mu U^{i_1}U^{i_2}...U^{i_k} \right)=0 \ .
\end{equation}
They give rise to the following hierarchy of conserved quantities:
\begin{equation}
\bK^{i_1...i_k} = \int d^{2p+1}x\, \cH^{(UBI)}_p\, U^{i_1}U^{i_2}...U^{i_k}\ .
\end{equation}
All these quantities are in involution, i.e.
\begin{equation}
\{ \bK^{i_1...i_k},\bK^{j_1...j_l}\}_p = 0\ .
\end{equation}
The only exception is $p=0$. In this case $|u^1|=1$ and all $\bK^{1...1}$
are aqual (up to a sign). But now the conformal group is infinite
dimensional and one has still an infinite number of constants of motion.

For other relation between Born-Infeld theory and fluid dynamics see e.g.
\cite{Jackiw}.

\section{Self-dual field}
\setcounter{equation}{0}

Note, that the  Maxwell eqs. for even $p$
 have the following form:
\beq           \label{dBM}
\partial_0 B^{I} &=&  \frac{1}{p!}\
\epsilon^{IkJ}\
\p_k D_{J}\ ,\\    \label{dDM}
\partial_0 D^{I} &=&  \frac{1}{p!}\
\epsilon^{IkJ}\
\p_k B_{J}\ ,
\eeq
and differ from the field eqs. of UBI (\ref{d1e})--(\ref{d2e})
 by the presence of the scaling parameter $|D|/|B|$.

Now, let us introduce chiral and anti-chiral combinations:
\beq
V^I_\pm = \frac{1}{\sqrt{2}} (D^I \pm B^I)\ .
\eeq
The Maxwell Hamiltonian rewritten in terms of $V_\pm$ has the following form
\begin{equation}
\cH_p^{Maxwell} = \frac{1}{2p!} \left(V_+\cdot V_+ + V_-\cdot V_-\right)\ ,
\end{equation}
and the Maxwell eqs. read:
\begin{equation}
V_\pm^I = \pm \frac{1}{p!} \epsilon^{IkJ}\p_k V_{J\pm}\ ,
\end{equation}
i.e. both components decouple and evolve independently. This property holds
for the Maxwell theory only. However, if the initial condition is such that
$V_-=0$, i.e. $D=B$ or $V_-=0$, i.e. $D=-B$, then for any $t$, $V_-(t)=0$ or
$V_+(t)=0$ also for the UBI theory defined by (\ref{d1e})-(\ref{d2e}). This
property holds for any $p$-form theory defined by a Hamiltonian satisfying
(\ref{O11}) and invariant under $Z_2\times Z_2$ given by
(\ref{Z1})-(\ref{Z2}). Therefore, for chiral (anti-chiral) data the dynamics
always reduces to
\begin{equation}   \label{pm}
B^I = \pm \frac{1}{p!} \epsilon^{IkJ}\p_k B_{J}\ ,
\end{equation}
i.e. it corresponds to (anti)self-dual field \cite{self}.
For (anti)self-dual data both Maxwell and UBI
Hamiltonians read:
\begin{equation}
\cH^{self}_p = \frac{1}{p!}\ B\cdot B\ ,
\end{equation}
and (\ref{pm}) defines the Hamiltonian system with respect to the following
canonical structures:
\begin{equation}
\{ B^I(\bx),B^J(\by)\} =
\pm \epsilon^{IkJ}\, \p_k\, \delta^{(2p+1)}(\bx-\by)\ .
\end{equation}

\section{Conclusions}
\setcounter{equation}{0}

Note, that $p$-form UBI is uniquely defined by the symmetry group. Namely, the
UBI Hamiltonian is a maximally symmetric solution of (\ref{O21}) or
(\ref{O11}) for odd and even $p$ respectively. Consider e.g. (\ref{O21}). Its
$SO(2,1)$-symmetric solution is a function of one variable $f=f(\tau)$ with
\beq
\tau = \alpha^2 -\beta^2 - \gamma^2\ . \nonumber
\eeq
One immediately shows that $f=\sqrt{\tau}$ in agreement with (\ref{HBI-1}).

For odd $p$ the $p$-form theory is duality invariant (i.e.
invariant under $SO(2)$ rotations (\ref{I})) iff \cite{Gibbons}
\begin{equation}   \label{Gib}
(\p_\eta L_p)^2 - (\p_\xi L_p)^2 =-1\ ,
\end{equation}
where
\beq
\eta &=& \sqrt{S^2_p + P^2_p}\ ,\nonumber\\
\xi &=& S_p\ . \nonumber
\eeq
Eq. (\ref{Gib}) has the same form as (\ref{O11}). Now, the maximally
symmetric solution to (\ref{Gib}) reads $L_p = |P_p|$, i.e. the corresponding
theory is trivial. Note, that performing $b\rightarrow 0$ limit in the BI
Lagrangian (\ref{LBI}) we  have also obtained $|P_p|$.

The  presence of an infinite hierarchy of constants of motion often implies
complete integrability of the theory. This question for $p=1$ UBI was posed in
\cite{IBB}. The answer is not known. It would be also interesting to find the
corresponding quantum version of this theory.

{\it Note added:} While this paper was being completed I received
unpublished notes \cite{Gib}, in which similar results were obtained.
Moreover, in \cite{Gib} the particles fluid of section \ref{FLUID} was
generalized to  $p$-brane fluids.

\section*{Acknowledgements} I thank prof. I. Bia{\l}ynicki-Birula for
pointing out the Ref. \cite{IBB} anf prof. G. W. Gibbons for sending me
his unpublished notes \cite{Gib}.
This work was partially supported by the KBN
Grant No. 2 P03A 047 15.

\end{document}